\documentclass[twocolumn,showpacs,preprintnumbers,amsmath,amssymb,endfloats*]{revtex4}
\usepackage{graphicx}

\begin{document}

\vspace*{1cm}
\thispagestyle{empty}
\title{
Comment on ``Lateral Casimir force beyond the proximity-force approximation''}

\author{F.~Chen${}^{1}$, U.~Mohideen${}^1$, G.~L.~Klimchitskaya${}^2$,
and V.~M.~Mostepanenko${}^2$}

\affiliation{${}^{1}$Department of Physics and Astronomy, 
University of California,
Riverside, California 92521, USA\\
${}^2$Institute for Theoretical Physics,
Leipzig University, Augustusplatz 10/11, 04109 Leipzig, Germany
}

\maketitle

The Letter \cite{1} is devoted to the calculation of the lateral Casimir force
arising between the corrugated metallic bodies (two parallel plates
or a sphere above a plate at a separation $L$) taking corrugations into 
account 
without use of the proximity force approximation (PFA). The metal of the test 
bodies is described by the plasma model ($\omega_P$ is the plasma frequency),
the corrugation wavelength is $\lambda_C$ and the amplitudes are $a=a_1,a_2$.
The lateral Casimir force with the amplitude of order of $a_1 a_2$
is found using the scattering approach. As the authors themselves note,
``This technique is valid as long as... $a\ll L, \lambda_C, \lambda_P$.''
Previously the lateral Casimir force was measured in 
Ref.~\cite{3}.
It was also calculated for test bodies made of ideal
metal outside the PFA \cite{4} and of real metals
described by the plasma model using the PFA \cite{3}. The main result of
Ref.~\cite{1} is that ``in realistic experimental situations the proximity
force approximation overestimates the force up to 30\%.'' Below we
demonstrate that the approach in Ref.~\cite{1} 
is in disagreement with a path-integral theory of Ref.~\cite{4}, 
and     
the approximations used are not appropriate for
comparison to experiment and for making statements on
the accuracy of the PFA.

We start with the computational results in Fig.1 of \cite{1} for the variation
of function $\rho$ versus $k=2\pi /\lambda_C$ describing the deviation from PFA
in the case of two parallel plates. According to this figure, 
with $\lambda_C=1.2\mu$m
at $L=200$nm the lateral force amplitude is less by 16\% than the value 
given by the
PFA. This is, however, in contradiction with a more fundamental theory 
\cite{4} 
formulated
for ideal metals. It is easily seen, that the quantity $\rho$, plotted in Fig.1
as a function of $k$ at different $L$, is, in fact, a function of $kL$. 
Thus, according to the Letter, for corrugated plates with the rescaled 
$\lambda_C=12\mu$m
and $L=2\mu$m the deviation of the lateral force amplitude from the 
PFA value is still
16\%. At $L=2\mu$m, however, the role of nonideality of a metal is
very small,
and Fig.5 of Ref.~\cite{4} demonstrates the complete agreement between the
exact result and the PFA if (as it holds in our case) $L$ is several 
times less than  $\lambda_C$. If this condition is not
satisfied, the PFA underestimates the force amplitude \cite{4}, and not
overestimates it as in \cite{1}. 
In the nonperturbative regime the same is demonstrated 
in Ref.~\cite{5}.

For the experimental configuration of a sphere at a separation
$L=221\,$nm above a plate
Ref.~\cite{1} obtains the ``exact'' computational value of 0.20 pN for the
amplitude of the lateral force with all experimental parameters 
as in Ref.~\cite{3}
($\lambda_C=1.2\mu$m, $\lambda_P=136\,$nm, $a_1=59\,$nm, and $a_2=8\,$nm).
According to Ref.~\cite{1}, the linear in $a_1 a_2$ version of the PFA gives 
instead 0.28 pN, i.e., 40\% difference. However, Ref.~\cite{1} does not
claim that the PFA overestimates the force by 40\%, but employs the
following reasoning: ``Precisely, Ref.~\cite{3} finds a
force of 0.32\,pN at $L=221\,$nm, with a relative correction due to higher
powers of 1.21. Discounting this factor, the second-order force should 
be 0.26\,pN, which overestimates the exact result by a factor of the order
of 30\%.''

At this point we stress that in Ref.~\cite{3} the force
amplitude at $L=221$ nm was both measured and computed 
using the complete PFA accounting for all powers in
$a_1, a_2$. The measured and theoretical values of 0.32$\pm 0.077\,$pN
(at 95\% confidence)
and 0.33\,pN, respectively, are in a very good agreement. It is illogical
that Ref.~\cite{1} ``discounts'' the effect due to higher powers in $a_1, a_2$
in our result if it aims to propose some general statements on the accuracy 
of the PFA. In our experiment 
$a_1$ and $a_2$ are not small
compared to $L$ (for instance, $a_1/L=0.27$) and it is insufficient to 
restrict to the first power in $a_1 a_2$ as is done in Ref.~\cite{1}. 
The calculations of Ref.~\cite{1} are also performed under the assumption
that $a_1,a_2\ll\lambda_P$ which is not met in our experiment because
$a_1/\lambda_P=0.43$. 
As a result, Ref.~\cite{1} arrives at a force amplitude of 0.20\,pN
so far away from the measured value of 0.32\,pN and theoretical 
value of 0.33\,pN
using the complete PFA \cite{3}. The 
force amplitude of 
0.20\,pN found in the Letter, when compared to experiment,
 could be realized only with the probability of 0.06\%. 
In the approach of Ref.~\cite{1} the force amplitude is underestimated by
30--40\% compared to both the PFA prediction and experiment.
This approach is inadequate, because the equations,
obtained to the lowest orders in small parameters, 
were applied outside of their application range, 
and the results obtained
are not in agreement with the path-integral theory of
Ref.~\cite{3}. Notice that the 
linear version of the PFA \cite{1}
leads to 0.28\,pN force amplitude instead of 0.20\,pN, i.e., much
closer to the measured and computed 
using the complete PFA \cite{3} values of 0.32\,pN
and 0.33\,pN, respectively. This
is in agreement with the result of Refs.~\cite{4,5} obtained for
ideal metal plates that the PFA works well when, as it holds in our
experiment, the separation $L$ is several times less than $\lambda_C$.

\end{document}